\documentclass[english]{article}
\usepackage[T1]{fontenc}
\usepackage[latin9]{inputenc}
\usepackage{geometry}
\usepackage{amsmath}
\usepackage{amssymb}
\usepackage{setspace}
\makeatletter

\providecommand{\tabularnewline}{\\}

\newcommand{\lyxaddress}[1]{
\par {\raggedright #1
\vspace{1.4em}
\noindent\par}
}

\makeatother

\usepackage{babel}
\begin{document}

\title{Octonion and Split Octonion Representation of $SO(8)$ Symmetry}

\author{Pushpa$^{(1)}$, P. S. Bisht$^{\left(1\right)}$, Tianjun Li$^{\left(2\right)}$
and O. P. S. Negi$^{\left(1\right)}$}

\maketitle

\lyxaddress{\begin{center}
$(1)$ Department of Physics,\\
 Kumaun University, S. S. J. Campus, \\
Almora-263601 (Uttarakhand) India\\

\par\end{center}}

\lyxaddress{\begin{center}
$(2)$ Institute of Theoretical Physics,\\
 Chinese Academy of Sciences,\\
 P. O. Box 2735, Beijing 100080, \\
P. R. China\\

\par\end{center}}

\lyxaddress{\begin{center}
Email-pushpakalauni60@yahoo.co.in\\
 ps\_bisht123@rediffmail.com \\
tli@itp.ac.cn,\\
ops\_negi@yahoo.co.in.
\par\end{center}}
\begin{abstract}
The 8 $\times$ 8 matrix representation of $SO(8)$ Symmetry has been
defined by using the direct product of Pauli matrices and Gamma matrices.
These 8 $\times$ 8 matrices are being used to describe the rotations
in $SO(8)$ symmetry. The comparison of 8$\times$8 matrices with
octonions has also been shown. The transformations of $SO(8)$ symmetry
are represented with the help of Octonions and split Octonions spinors.
\end{abstract}

\section{Introduction}

Octonions form a division algebra of the highest possible dimension
8. The group of rotations in eight dimensions has been described \cite{key-1,key-2}
as the extensions for the symmetries of elementary particles. G$\ddot{u}$
naydin and G$\ddot{u}$rsey \cite{key-3} discussed the Lie algebra
of $G_{2}$ group and its embedding \cite{key-4,key-5,key-6} in $SO(7)$
and $SO(8)$ groups. The triality properties of the rotation group
$SO(8)$ which are closely related to octonions are described by Curtright
\cite{key-7}. A dynamical scheme of quark and lepton family unification
based on non associative algebra has also been discussed \cite{key-8}.
Generators of $SO(8)$ are constructed by using Octonion structure
tensors \cite{key-9}, and the representations of these generators
are given as products of Octonions. Lassig and Joshi \cite{key-10}
introduced the bi-modular representation of octonions and formulated
the $SO(8)$ gauge theory equivalent to the octonionic construction.\textbf{
}Furthermore, some attention has been given to octonions \cite{key-11}
in theoretical physics in order to extend the 3+1 space-time to eight
dimensional space-time as the consequence to accommodate the ever
increasing quantum numbers and internal symmetries related to elementary
particles and gauge fields. It is shown \cite{key-12} that three
dimensional vector space may be extended to seven dimensional space
by means of octonions under certain permutations of combinations of
structure constant associated with the octonion multiplication rules.
Recently, we have used quaternions and octonions to defined the Quantum
Chromo Dynamics \cite{key-13}, symmetry breaking \cite{key-14},
Flavor symmetry \cite{key-15} and Casimir operetor \cite{key-16}
in their algebraic form. Here, we have constructed the 8$\times$8
matrix by using Pauli matrices and Gamma matrices which represents
the $SO(8)$ symmetry. The eight dimensional space has been defined
by considering octonions as a spinor.

\section{Matrix representation of $SO(8)$ Symmetry}

$SO(8)$ represents the special orthogonal group of eight-dimensional
rotations. By using direct product of Pauli matrices and Gamma matrices,
we have constructed the eight dimensional representation of $SO(8)$
Symmetry.\\
As the Gamma matrices for Dirac Pauli representation are given by
\begin{align}
\gamma_{j}=\left[\begin{array}{cc}
0 & -i\sigma_{j}\\
i\sigma_{j} & 0
\end{array}\right],\,\,\,\, & \gamma_{4}=\left[\begin{array}{cc}
I & 0\\
0 & -I
\end{array}\right]\label{eq:1}
\end{align}
where $\sigma_{j}$ are Pauli matrices and $I$ is a 2 $\times$ 2
unit matrix.\\
Now to construct the eight dimensional representation \cite{key-3}
of $SO(8)$ Symmetry, we have chosen the following representations,

\begin{align}
\beta_{1}=\sigma_{1}\otimes\gamma_{1}= & i\left[\Sigma_{36}+\Sigma_{45}+\Sigma_{72}+\Sigma_{81}-\Sigma_{18}-\Sigma_{27}-\Sigma_{54}-\Sigma_{63}\right];\nonumber \\
\beta_{2}=\sigma_{3}\otimes\gamma_{1}= & i\left[\Sigma_{32}+\Sigma_{41}+\Sigma_{58}+\Sigma_{67}-\Sigma_{14}-\Sigma_{23}-\Sigma_{76}-\Sigma_{85}\right];\nonumber \\
\beta_{3}=\sigma_{2}\otimes\gamma_{3}= & \left[\Sigma_{28}+\Sigma_{82}+\Sigma_{35}+\Sigma_{53}-\Sigma_{64}-\Sigma_{46}-\Sigma_{71}-\Sigma_{17}\right];\nonumber \\
\beta_{4}=\sigma_{3}\otimes\gamma_{2}= & \left[\Sigma_{23}+\Sigma_{32}+\Sigma_{58}+\Sigma_{85}-\Sigma_{14}-\Sigma_{41}-\Sigma_{67}-\Sigma_{76}\right];\nonumber \\
\beta_{5}=\sigma_{1}\otimes\gamma_{3}= & i\left[\Sigma_{28}+\Sigma_{35}+\Sigma_{64}+\Sigma_{71}-\Sigma_{17}-\Sigma_{46}-\Sigma_{53}-\Sigma_{82}\right];\nonumber \\
\beta_{6}=\sigma_{3}\otimes\gamma_{3}= & i\left[\Sigma_{24}+\Sigma_{31}+\Sigma_{57}+\Sigma_{86}-\Sigma_{13}-\Sigma_{42}-\Sigma_{68}-\Sigma_{75}\right];\nonumber \\
\beta_{7}=\sigma_{1}\otimes\gamma_{4}= & \left[\Sigma_{15}+\Sigma_{26}+\Sigma_{51}+\Sigma_{62}-\Sigma_{37}-\Sigma_{48}-\Sigma_{73}-\Sigma_{84}\right];\nonumber \\
\beta_{8}=\sigma_{1}\otimes\gamma_{1}= & \left[\Sigma_{11}+\Sigma_{22}+\Sigma_{77}+\Sigma_{88}-\Sigma_{33}-\Sigma_{44}-\Sigma_{55}-\Sigma_{66}\right];\label{eq:2}
\end{align}
where $\otimes$ denotes the direct product of matrices. $\sigma^{\shortmid}$s
are Pauli matrices and $\gamma^{\shortmid}$s are Dirac matrices.
These 8 matrices $\beta_{1}$, $\beta_{2}$, .........., $\beta_{8}$
are 8 $\times$ 8 Hermitian matrices and $\Sigma_{mn}$ are 8 $\times$
8 matrices in which $mn^{th}$ matrix element is unity and rest elements
are zero. $\Sigma_{mn}$ are the 8 $\times$ 8 matrix representation
of the generators $SO(8)$. \\
The eight dimensional space on which $SO(8)$ acts can be given by
the structure of a non associative algebra. Here such algebra is described
by the octonions as a general spinor $\psi$ in eight dimensions.
Under this symmetry, the spinor $\psi$ transforms as 

\begin{align}
\psi\longmapsto\psi^{\shortmid}= & \exp\left[\sum_{A=1}^{8}f_{A}\beta_{A}\right]\psi\nonumber \\
= & e^{X}\psi\label{eq:3}
\end{align}
where vector $X$ is, 

\begin{align}
X= & \sum_{A=1}^{8}f_{A}\beta_{A}\label{eq:4}
\end{align}
with $f_{1}$, $f_{2}$, $f_{3}$,..........$f_{8}$ as the component
of vector. Since the representation given in equation $(\ref{eq:2})$
will contain a vector space of dimension 8, for which we want to introduce
an octonionic description. This description, should be invariant under
the appropriate $SO(8)$ symmetry group. Therefore Spinor $\psi$
is defined as, 
\begin{align}
\psi= & \left[\begin{array}{c}
1\\
e_{1}\\
e_{2}\\
e_{3}\\
e_{4}\\
e_{5}\\
e_{6}\\
e_{7}
\end{array}\right]\label{eq:5}
\end{align}
where $e_{1}$, $e_{2}$,...........$e_{7}$ are Octonion units. After
expanding equation $\left(\ref{eq:4}\right)$, X becomes as,
\begin{align}
X= & \left[\begin{array}{cccccccc}
f_{8} & 0 & -if_{6} & -f_{4}-if_{2} & f_{7} & 0 & -f_{3}-if_{5} & -if_{1}\\
0 & f_{8} & f_{4}-if_{2} & if_{6} & 0 & f_{7} & -if_{1} & f_{3}+if_{5}\\
if_{6} & f_{4}+if_{2} & -f_{8} & 0 & f_{3}+if_{5} & if_{1} & -f_{7} & 0\\
-f_{4}+if_{2} & -if_{6} & 0 & -f_{8} & if_{1} & -f_{3}-if_{5} & 0 & -f_{7}\\
f_{7} & 0 & f_{3}-if_{5} & -if_{1} & -f_{8} & 0 & if_{6} & f_{4}+if_{2}\\
0 & f_{7} & -if_{1} & -f_{3}+if_{5} & 0 & -f_{8} & -f_{4}+if_{2} & -if_{6}\\
-f_{3}+if_{5} & if_{1} & -f_{7} & 0 & -if_{6} & -f_{4}-if_{2} & f_{8} & 0\\
if_{1} & f_{3}-if_{5} & 0 & -f_{7} & f_{4}-if_{2} & if_{6} & 0 & f_{8}
\end{array}\right]\label{eq:6}
\end{align}
which is a traceless Hermitian matrix. In compact form X can be written
as,

\begin{align}
X= & \left[\begin{array}{cc}
A & B^{\dagger}\\
B & -A
\end{array}\right]\label{eq:7}
\end{align}
where $A$ and $B$ are given as, 
\begin{gather}
A=\left[\begin{array}{cccc}
f_{8} & 0 & -if_{6} & -f_{4}-if_{2}\\
0 & f_{8} & f_{4}-if_{2} & if_{6}\\
if_{6} & f_{4}+if_{2} & -f_{8} & 0\\
-f_{4}+if_{2} & -if_{6} & 0 & -f_{8}
\end{array}\right];\label{eq:8}\\
B=\left[\begin{array}{cccc}
f_{7} & 0 & f_{3}-if_{5} & -if_{1}\\
0 & f_{7} & -if_{1} & -f_{3}+if_{5}\\
-f_{3}+if_{5} & if_{1} & -f_{7} & 0\\
if_{1} & f_{3}-if_{5} & 0 & -f_{7}
\end{array}\right].\label{eq:9}
\end{gather}
Matrices A and B are independent of each other.\\
Furthermore, the constructed 8$\times$8 matrices given in equation
$\left(\ref{eq:2}\right)$ are being used to describe the rotation
in $SO(8)$ Symmetry.

\section{Rotation in SO(8) Representation}

As an infinitesimal rotation by an angle $\theta$ in the plane ($k$,
$l$) is obtained by the following operator \cite{key-1},

\begin{align}
R_{kl}= & 1+\theta\beta_{k}\beta_{l}\label{eq:10}
\end{align}
which acts on a vector $X$ to form a rotated vector $X^{\shortmid}$
as,

\begin{align}
X^{\shortmid}= & R_{kl}XR_{kl}^{-1}\label{eq:11}
\end{align}
By using Equation $\left(\ref{eq:10}\right)$, the rotation operator
$R_{12}$ becomes,

\begin{align}
R_{12}= & \left[\begin{array}{cccccccc}
1 & 0 & 0 & 0 & -\theta & 0 & 0 & 0\\
0 & 1 & 0 & 0 & 0 & -\theta & 0 & 0\\
0 & 0 & 1 & 0 & 0 & 0 & -\theta & 0\\
0 & 0 & 0 & 1 & 0 & 0 & 0 & -\theta\\
\theta & 0 & 0 & 0 & 1 & 0 & 0 & 0\\
0 & \theta & 0 & 0 & 0 & 1 & 0 & 0\\
0 & 0 & \theta & 0 & 0 & 0 & 1 & 0\\
0 & 0 & 0 & \theta & 0 & 0 & 0 & 1
\end{array}\right].\label{eq:12}
\end{align}
The rotation $R_{12}$ gives the rotated vector $X^{\shortmid}$ as
follows, 

\begin{align}
X^{\shortmid}= & X+2\theta\left[\begin{array}{cccccccc}
-f_{7} & 0 & if_{5} & if_{1} & f_{8} & 0 & -if_{6} & -(f_{4}+if_{2})\\
0 & -f_{7} & if_{1} & -if_{5} & 0 & f_{8} & (f_{4}-if_{2}) & if_{6}\\
-if_{5} & -if_{1} & f_{7} & 0 & if_{6} & f_{4}+if_{2} & -f_{8} & 0\\
-if_{1} & if_{5} & 0 & f_{7} & -f_{4}+if_{2} & -if_{6} & 0 & -f_{8}\\
f_{8} & 0 & -if_{6} & -(f_{4}+if_{2}) & f_{7} & 0 & -if_{5} & -if_{1}\\
0 & f_{8} & (f_{4}-if_{2}) & if_{6} & 0 & f_{7} & -if_{1} & if_{5}\\
if_{6} & \left(f_{4}+if_{2}\right) & -f_{8} & 0 & if_{5} & if_{1} & -f_{7} & 0\\
-(f_{4}-if_{2}) & -if_{6} & 0 & -f_{8} & if_{1} & -if_{5} & 0 & -f_{7}
\end{array}\right].\label{eq:13}
\end{align}
Since $\theta$ is an infinitesimal rotation, therefore neglecting
$\theta^{2}$ terms. Rotation $R_{12}$ transforms the components
of a vector $X$ as follows:

\begin{align}
f_{1}\Longrightarrow & f_{1}-2\theta i\left(f_{4}+if_{2}\right);\nonumber \\
f_{2}\Longrightarrow & f_{2}-2\theta f_{1};\nonumber \\
f_{3}\Longrightarrow & f_{3}+2\theta if_{6};\nonumber \\
f_{4}\Longrightarrow & f_{4}-2\theta if_{1};\nonumber \\
f_{5}\Longrightarrow & f_{5}+2\theta f_{6};\nonumber \\
f_{6}\Longrightarrow & f_{6}-2\theta f_{5};\nonumber \\
f_{7}\Longrightarrow & f_{7}+2\theta f_{8};\nonumber \\
f_{8}\Longrightarrow & f_{8}-2\theta f_{7}.\label{eq:14}
\end{align}
We have calculated all the possible rotations of $R_{kl}$, and observed
that $R_{56}$ and $R_{78}$ give the same transformation as $R_{12}$.

\section{Comparison of 8$\times$8 matrix with Octonions}

As the Octonions are described over the algebra of real numbers having
the vector space of dimension $8$. Here, we construct following matrices
using $\beta$ matrices for showing similarity between eight dimensional
matrices with Octonion basis elements. \\
Let 
\begin{align}
E_{0}= & I_{8}\nonumber \\
E_{1}\Longrightarrow & \beta_{1}\beta_{5}=\beta_{2}\beta_{6};\nonumber \\
E_{2}\Longrightarrow & \beta_{1}\beta_{7}=\beta_{2}\beta_{8};\nonumber \\
E_{3}\Longrightarrow & \beta_{7}\beta_{5}=\beta_{8}\beta_{6};\nonumber \\
E_{4}\Longrightarrow & \beta_{7};\nonumber \\
E_{5}\Longrightarrow & \beta_{5};\nonumber \\
E_{6}\Longrightarrow & \beta_{1};\nonumber \\
E_{7}\Longrightarrow & \beta_{7}\beta_{5}\beta_{1}=\beta_{8}\beta_{6}\beta_{1}=E_{3}E_{6}.\label{eq:15}
\end{align}
where $I_{8}$ is 8 $\times$8 unity matrix. Multiplication Table
for the above matrices are given as in Table 1.

\begin{table}
\begin{centering}
\begin{tabular}{|c|c|c|c|c|c|c|c|c|}
\hline 
. & $E_{0}$ & $E_{1}$ & $E_{2}$ & $E_{3}$ & $E_{4}$ & $E_{5}$ & $E_{6}$ & $E_{7}$\tabularnewline
\hline 
$E_{0}$ & $E_{0}$ & $E_{1}$ & $E_{2}$ & $E_{3}$ & $E_{4}$ & $E_{5}$ & $E_{6}$ & $E_{7}$\tabularnewline
\hline 
$E_{1}$ & $E_{1}$ & -$E_{0}$ & $E_{3}$ & -$E_{2}$ & -$E_{7}$ & $E_{6}$ & -$E_{5}$ & $E_{4}$\tabularnewline
\hline 
$E_{2}$ & $E_{2}$ & $-E_{3}$ & -$E_{0}$ & $E_{1}$ & $E_{6}$ & $E_{7}$ & -$E_{4}$ & -$E_{5}$\tabularnewline
\hline 
$E_{3}$ & $E_{3}$ & $E_{2}$ & -$E_{1}$ & -$E_{0}$ & -$E_{5}$ & $E_{4}$ & $E_{7}$ & -$E_{6}$\tabularnewline
\hline 
$E_{4}$ & $E_{4}$ & -$E_{7}$ & -$E_{6}$ & $E_{5}$ & $E_{0}$ & $E_{3}$ & -$E_{2}$ & -$E_{1}$\tabularnewline
\hline 
$E_{5}$ & $E_{5}$ & -$E_{6}$ & $E_{7}$ & -$E_{4}$ & -$E_{3}$ & $E_{0}$ & -$E_{1}$ & $E_{2}$\tabularnewline
\hline 
$E_{6}$ & $E_{6}$ & $E_{5}$ & $E_{4}$ & $E_{7}$ & $E_{2}$ & -$E_{1}$ & $E_{0}$ & $E_{3}$\tabularnewline
\hline 
$E_{7}$ & $E_{7}$ & $E_{4}$ & -$E_{5}$ & -$E_{6}$ & -$E_{1}$ & $E_{2}$ & $E_{3}$ & -$E_{0}$\tabularnewline
\hline 
\end{tabular}
\par\end{centering}

\caption{8$\times8$ matrix multiplication}
\end{table}
Octonion multiplication\textbf{ \cite{key-8}} are given in Table
2.

\begin{table}
\begin{centering}
\begin{tabular}{|c|c|c|c|c|c|c|c|c|}
\hline 
. & $e_{0}$ & $e_{1}$ & $e_{2}$ & $e_{3}$ & $e_{4}$ & $e_{5}$ & $e_{6}$ & $e_{7}$\tabularnewline
\hline 
$e_{0}$ & $e_{0}$ & $e_{1}$ & $e_{2}$ & $e_{3}$ & $e_{4}$ & $e_{5}$ & $e_{6}$ & $e_{7}$\tabularnewline
\hline 
$e_{1}$ & $e_{1}$ & -$e_{0}$ & $e_{3}$ & -$e_{2}$ & $e_{7}$ & -$e_{6}$ & $e_{5}$ & -$e_{4}$\tabularnewline
\hline 
$e_{2}$ & $e_{2}$ & -$e_{3}$ & -$e_{0}$ & $e_{1}$ & $e_{6}$ & $e_{7}$ & -$e_{4}$ & -$e_{5}$\tabularnewline
\hline 
$e_{3}$ & $e_{3}$ & $e_{2}$ & -$e_{1}$ & -$e_{0}$ & -$e_{5}$ & $e_{4}$ & $e_{7}$ & -$e_{6}$\tabularnewline
\hline 
$e_{4}$ & $e_{4}$ & -$e_{7}$ & -$e_{6}$ & $e_{5}$ & -$e_{0}$ & -$e_{3}$ & $e_{2}$ & $e_{1}$\tabularnewline
\hline 
$e_{5}$ & $e_{5}$ & $e_{6}$ & -$e_{7}$ & -$e_{4}$ & $e_{3}$ & -$e_{0}$ & -$e_{1}$ & $e_{2}$\tabularnewline
\hline 
$e_{6}$ & $e_{6}$ & -$e_{5}$ & $e_{4}$ & -$e_{7}$ & -$e_{2}$ & $e_{1}$ & -$e_{0}$ & $e_{3}$\tabularnewline
\hline 
$e_{7}$ & $e_{7}$ & $e_{4}$ & $e_{5}$ & $e_{6}$ & -$e_{1}$ & -$e_{2}$ & -$e_{3}$ & -$e_{0}$\tabularnewline
\hline 
\end{tabular}
\par\end{centering}

\caption{Octonion multiplication}
\end{table}
On comparing 8$\times$8 matrix multiplication table (Table 1) with
the octonion multiplication table (Table 2), we can see some similarities
between these two. It means that multiplication of 8$\times$8 matrices
satisfied some of the multiplication of the 8 dimensional algebra
of Octonions. Out of 64 combinations, 48 combinations are identical,
however, rest 16 combinations have opposite signs for octonions and
8$\times$8 matrix.\textbf{ }This dissimilarity can be attributed
to the non associativity of octonions.\textbf{ }Thus, we have made
a comparative study between the multiplicative properties of the 8$\times$8
matrix and the octonions.

\section{Split Octonions }

Split Octonion algebra \cite{key-17} with its split base units is
defined as 

\begin{gather}
u_{0}=\frac{1}{2}\left(e_{0}+ie_{7}\right),\,\,\,\,\,\, u_{0}^{\star}=\frac{1}{2}\left(e_{0}-ie_{7}\right);\label{eq:16}\\
u_{m}=\frac{1}{2}\left(e_{m}+ie_{m+3}\right)\,\,\,\,\, u_{m}^{\star}=\frac{1}{2}\left(e_{m}-ie_{m+3}\right).\label{eq:17}
\end{gather}
where m=1,2,3. \\
These basis element satisfy the following algebra

\begin{align}
u_{i}u_{j}=-u_{j}u_{i}=\epsilon_{ijk}u_{k}^{\star},\,\,\,\,\,\, & u_{i}^{\star}u_{j}^{\star}=-u_{j}^{\star}u_{i}^{\star}=\epsilon_{ijk}u_{k};\nonumber \\
u_{i}u_{j}^{\star}=-\delta_{ij}u_{0},\,\,\,\,\,\,\,\,\,\,\,\,\,\,\,\, & u_{i}^{\star}u_{j}=-\delta_{ij}u_{0}^{\star};\nonumber \\
u_{0}u_{i}=u_{i}u_{0}^{\star}=u_{i},\,\,\,\,\,\,\,\,\,\,\,\,\, & u_{0}^{\star}u_{i}^{\star}=u_{i}^{\star}u_{0}=u_{i}^{\star};\nonumber \\
u_{i}u_{0}=u_{0}u_{i}^{\star}=0,\,\,\,\,\,\,\,\,\,\,\,\,\, & u_{i}^{\star}u_{0}^{\star}=u_{0}^{\star}u_{i}=0;\nonumber \\
u_{0}u_{0}^{\star}=u_{0}^{\star}u_{0}=0\,\,\,\,\,\,\,\,\,\,\,\,\, & u_{0}^{2}=u_{0},\,\, u_{0}^{\star2}=u_{0}^{\star}.\label{eq:18}
\end{align}
These relations $\left(\ref{eq:18}\right)$ are invariant \cite{key-18}
under $G_{2}$ group as a automorphism of octonions. Unlike octonions,
the split octonion algebra contains zero divisors and is therefore
not a division algebra.

\section{Split Octonion representation of SO(8) Symmetry}

Now, spinor $\psi$ in terms of split octonion \cite{key-3} i.e.

\begin{align}
\phi= & \left[\begin{array}{c}
u_{0}\\
u_{1}\\
u_{2}\\
u_{3}\\
u_{0}^{\star}\\
u_{1}^{\star}\\
u_{2}^{\star}\\
u_{3}^{\star}
\end{array}\right]=\left[\begin{array}{c}
\frac{1}{2}\left(1+ie_{7}\right)\\
\frac{1}{2}\left(e_{1}+ie_{4}\right)\\
\frac{1}{2}\left(e_{2}+ie_{5}\right)\\
\frac{1}{2}\left(e_{3}+ie_{6}\right)\\
\frac{1}{2}\left(1-ie_{7}\right)\\
\frac{1}{2}\left(e_{1}-ie_{4}\right)\\
\frac{1}{2}\left(e_{2}-ie_{5}\right)\\
\frac{1}{2}\left(e_{3}-ie_{6}\right)
\end{array}\right]=\left[\begin{array}{c}
u\\
u^{\star}
\end{array}\right].\label{eq:19}
\end{align}
The mapping due to split octonions can be given as

\begin{align}
\phi\longmapsto\phi^{\shortmid} & =e^{Y}\phi\label{eq:20}
\end{align}
where
\begin{gather}
Y=\left[\begin{array}{cccccccc}
f_{8} & 0 & -if_{6} & -f_{4}-if_{2} & f_{8} & 0 & -if_{6} & -f_{4}-if_{2}\\
0 & f_{8} & f_{4}-if_{2} & if_{6} & 0 & f_{8} & f_{4}-if_{2} & if_{6}\\
if_{6} & f_{4}+if_{2} & -f_{8} & 0 & if_{6} & f_{4}+if_{2} & -f_{8} & 0\\
-f_{4}+if_{2} & -if_{6} & 0 & -f_{8} & -f_{4}+if_{2} & -if_{6} & 0 & -f_{8}\\
f_{7} & 0 & f_{3}-if_{5} & -if_{1} & f_{7} & 0 & f_{3}-if_{5} & -if_{1}\\
0 & f_{7} & -if_{1} & -f_{3}+if_{5} & 0 & f_{7} & -if_{1} & -f_{3}+if_{5}\\
-f_{3}+if_{5} & if_{1} & -f_{7} & 0 & -f_{3}+if_{5} & if_{1} & -f_{7} & 0\\
if_{1} & f_{3}-if_{5} & 0 & -f_{7} & if_{1} & f_{3}-if_{5} & 0 & -f_{7}
\end{array}\right]+\nonumber \\
\left[\begin{array}{cccccccc}
-f_{1} & -if_{7} & 0 & -f_{5}+if_{3} & f_{1} & if_{7} & 0 & -if_{3}+f_{5}\\
f_{5}-if_{3} & 0 & -if_{7} & -f_{1} & -f_{5}+if_{3} & 0 & if_{7} & f_{1}\\
0 & f_{5}-if_{3} & f_{1} & if_{7} & 0 & if_{3}-f_{5} & -f_{1} & -if_{7}\\
if_{7} & f_{1} & -f_{5}+if_{3} & 0 & -if_{7} & -f_{1} & f_{5}-if_{3} & -f_{8}\\
-i(f_{4}+if_{2}) & if_{8} & 0 & f_{6} & i(f_{4}+if_{2}) & -if_{8} & 0 & -f_{6}\\
-f_{6} & 0 & if_{8} & if_{4}-f_{2} & f_{6} & 0 & -if_{8} & -if_{4}+f_{2}\\
0 & -f_{6} & if_{4}-f_{2} & f_{8} & 0 & f_{6} & -if_{4}+f_{2} & -f_{8}\\
-if_{8} & -if_{4}-f_{2} & -if_{6} & 0 & if_{8} & if_{4}+f_{2} & if_{6} & 0
\end{array}\right].\label{eq:21}
\end{gather}

\begin{align}
Y= & \left[\begin{array}{cc}
A & A\\
B & B
\end{array}\right]+\left[\begin{array}{cc}
C & -C\\
D & -D
\end{array}\right]\nonumber \\
= & \left[\begin{array}{cc}
A+B & A-C\\
B+D & B-D
\end{array}\right]\label{eq:22}
\end{align}
Y is calculated by using equation $\left(\ref{eq:5}\right)$, $\left(\ref{eq:6}\right)$
$\left(\ref{eq:16}\right)$ and equation $\left(\ref{eq:17}\right)$.
$A$ and $B$ are already defined in equations $\left(\ref{eq:8}\right)$
and $\left(\ref{eq:9}\right)$. $C$ and $D$ are 4$\times$4 matrices
given as,

\begin{align}
C & =\left[\begin{array}{cccc}
-f_{1} & -if_{7} & 0 & -f_{5}+if_{3}\\
f_{5}-if_{3} & 0 & -if_{7} & -f_{1}\\
0 & f_{5}-if_{3} & f_{1} & if_{7}\\
if_{7} & f_{1} & -f_{5}+if_{3} & 0
\end{array}\right]\label{eq:23}
\end{align}
and 

\begin{align}
D= & \left[\begin{array}{cccc}
-i(f_{4}+if_{2}) & if_{8} & 0 & f_{6}\\
-f_{6} & 0 & if_{8} & if_{4}-f_{2}\\
0 & -f_{6} & if_{4}-f_{2} & f_{8}\\
-if_{8} & -if_{4}-f_{2} & -if_{6} & 0
\end{array}\right].\label{eq:24}
\end{align}
Equation (\ref{eq:22}) reprent how the SO(8) Symmetry split the Octonion
reprentation given in equation $(\ref{eq:7})$. This is the split
octonion representation in $SO(8)$ Symmetry.

\section{Result and Discussion}

The mathematical properties on the space of eight dimensions are presented
for their possible applications for the study of symmetries of elementary
particles. Here the eight dimensional space on which $SO(8)$ acts
has been defined in terms 8$\times$8 matrices. These 8$\times$8
matrix transformations are used as a general spinor in eight dimensions.
An infinitesimal rotation transformations in 8$\times$8 matrices
are defined. After using the different combinations of these matrices,
we compare the similarity of octonions with these matrices. Split
octonion representation of $SO(8)$ symmetry and their transformations
are also defined. These representations of 8 dimensional orthogonal
groups may be used to give octonionic descriptions of the Clifford
groups. Since the generators of our fundamental representation are
also generators of the $SO(8)$ symmetry group. Therefore, we have
used the representation of the octonion algebra of the $SO(8)$ symmetry
group. Octonionic symmetry has been used to represents the $SO(8)$
symmetry. We have calculated an embedding of the octonion symmetry
in $SO(8)$. A similar description for the exceptional Lie group $G_{2}$,
which is the automorphism group of the octonionic algebra has also
been found earlier \cite{key-3}. The octonionic description of the
vector representations of $SO(8)$ can give a unified picture of the
triality automorphisms of $SO(8)$.\\
\\
\textbf{ACKNOWLEDGEMENT}: One of us (Pushpa) is thankful to Dr. Ashok
Kumar for his careful reading of the manuscript and suggesting corrections.
Also she acknowledges Physical Research Laboratory, Ahmedabad for
allowing her to access the institute library for literature survey.

\end{document}